# Regge behaviour of structure functions and DGLAP evolution equation in leading order

U. Jamil[1] and J.K. Sarma[2]

Physics Department, Tezpur University, Napaam, Tezpur-784028, Assam, India

**Abstract.** In this paper, we find the t and x-evolution of deuteron structure function $F_2^d$ from DGLAP evolution equation of singlet structure function in leading order at small-x assuming the Regge behaviour of the singlet structure function at this limit and we compare our result of deuteron structure function with New Muon Collaboration data to find the range of the intercept $\lambda_d$ of Regge behaviour for deuteron structure function. We also discuss the limitations of Taylor series expansion method in the Regge behaviour of structure function.

## 1 Introduction

The measurements of the $F_2(x, Q^2)$ (proton, neutron and deuteron) structure functions by deep inelastic scattering (DIS) processes in the small-x region [1], where x is the Bjorken variable meaning fractional momentum carried by each parton i.e., quarks and gluons, and $Q^2$ is the four momentum of the exchanged gauge boson, have opened a new era in parton density measurements inside hadrons. The structure function reflects the momentum distributions of the partons in the nucleon. Because the proton and the neutron have different contents of up and down quarks, a measurement of $F_2^P$ and $F_2^d$ structure functions, together with isospin symmetry, provides a constraint on the individual quark distributions. This is an important phenomenological input to the calculation of many strong interaction processes. In addition, the $Q^2$ and x dependence of structure function can be used to test perturbative quantum chromodynamics (PQCD) [2, 3]. In PQCD, the high-$Q^2$ behaviour of DIS is given by the Dokshitzer-Gribov-Lipatov-Altarelli-Parisi (DGLAP) evolution equations [4]. In the double asymptotic limit (large energies, i.e. small-x and large photon virtualities $Q^2$), the DGLAP evolution equations [5] can be solved [6] and structure function is expected to rise approximately like a power of x towards small-x. Accordingly the approximate solutions of DGLAP evolution equations are reported in recent years [7-11] with considerable phenomenological success.

The small-x region of DIS offers a unique possibility to explore the Regge limit of PQCD. Two-body scattering of hadrons by leptons is strongly dominated by small momentum transfer or equivalently by small scattering angle. And the high energy limit, when the scattering energy is kept much greater than the external masses (and momentum transfer), is, by definition, the Regge limit. In DIS, $Q^2$ is, by definition, also kept large i.e. $Q^2 \gg \Lambda^2$, $\Lambda$ is the QCD scale parameter. The limit of large ν and $2M\nu \gg Q^2$ is therefore the Regge limit of DIS, where ν correspond to the virtuality of the exchanged boson and its energy in the nucleon rest frame and x is finite. The fact that $Q^2$ is large allows to use PQCD [12, 13]. This theory is successfully described by the exchange of a particle with appropriate quantum numbers and the exchange particle is called a Regge poles. The Regge poles carrying the quantum numbers of the vacuum and describing diffractive scattering is called pomeron which can be thought of as corresponding to an exchange of a pair of gluons. Other Regge poles are called reggeons and can be thought of as corresponding to the exchange of quarks and gluons. The Regge behaviour of the structure function $F_2$ in the large-$Q^2$ region reflects itself in the small-x behaviour of the quark and antiquark distributions [13-15]. Thus the Regge behaviour of the sea quark and antiquark distributions for small-x is given by $q_{sea}(x) \sim x^{\alpha p}$ with pomeron exchange of intercept $\alpha p = -1$. But the valence quark distribution for small-x given by $q_{val}(x) \sim x^{-\alpha r}$ corresponds to a reggeon exchange of intercept $\alpha r = 1/2$.

---

[1]jamil@tezu.ernet.in     [2]jks@tezu.ernet.in

In our present work, we have derived the solutions of singlet and non-singlet DGLAP evolution equations in leading order (LO) at small-x limit applying Regge theory. The LO deuteron structure functions results for t and x-evolutions are compared with New Muon Collaboration (NMC) small-x ($0.0045 \leq x \leq 0.0175$) and medium-$Q^2$ ($0.75 \leq Q^2 \leq 7$ GeV$^2$) data [16,17]. Here we overcome the limitations arose from Taylor series expansion method and this method is also mathematically simple. In this paper, section 1, section 2, section 3 and section 4 are the introduction, theory, results and discussion and conclusions respectively.

**2 Theory**

In the LO analysis, deuteron structure function is directly related to singlet structure function $F_2^S(x, Q^2)$ [18]. On the other hand, the differential coefficient of singlet structure function $F_2^S(x, Q^2)$ with respect to $\ln Q^2$ i.e. $\partial F_2^S(x, Q^2)/\partial \ln Q$ has a relation with singlet structure function itself as well as gluon distribution function from DGLAP evolution equations [19-21]. The LO DGLAP evolution equations for singlet and non-singlet structure functions [18] have the standard forms

$$\frac{\partial F_2^S(x,t)}{\partial t} - \frac{A_f}{t}\left[\{3+4\ln(1-x)\}F_2^S(x,t) + 2\int_x^1 \frac{d\omega}{1-\omega}\{(1+\omega^2)F_2^S(x/\omega,t) - 2F_2^S(x,t)\}\right.$$
$$\left. + \frac{3}{2}N_f \int_x^1 \{\omega^2 + (1-\omega)^2\}G(x/\omega,t)d\omega\right] = 0, \qquad (1)$$

$$\frac{\partial F_2^{NS}(x,t)}{\partial t} - \frac{A_f}{t}\left[\{3+4\ln(1-x)\}F_2^{NS}(x,t) + 2\int_x^1 \frac{d\omega}{1-\omega}\{(1+\omega^2)F_2^{NS}(x/\omega,t) - 2F_2^{NS}(x,t)\}\right] = 0, \quad (2)$$

where $t = \ln(Q^2/\Lambda^2)$ and $A_f = 4/(33 - 2N_f)$, $N_f$ being the number of flavours and $\Lambda$ is the QCD cut-off parameter.

Among the various methods to solve these equations, one important and simple method is to use Taylor expansion [22] to transform the integro-differential equations into partial differential equations and thus to solve them by standard methods [23, 24]. But when we apply Regge theory for the evolution equations and consider Regge behaviour of structure functions, the use of Taylor expansion becomes limited. In this method, we introduce the variable $u = 1-\omega$ and we get

$$\frac{x}{\omega} = \frac{x}{1-u} = x\sum_{k=0}^{\infty} u^k .$$

Since $0 < u < (1-x)$, where $|u| < 1$, $\frac{x}{1-u} = x\sum_{k=0}^{\infty} u^k$ is convergent. Applying the Taylor expansion for the singlet structure function in equation (1), we get

$$F_2^S\left(\frac{x}{\omega}, t\right) = F_2^S\left(\frac{x}{1-u}, t\right) = F_2^S\left(x + x\sum_{k=1}^{\infty} u^k, t\right)$$

$$= F_2^S(x,t) + x\sum_{k=1}^{\infty} u^k \frac{\partial F_2^S(x,t)}{\partial x} + \frac{1}{2} x^2 \left(\sum_{k=1}^{\infty} u^k\right)^2 \frac{\partial^2 F_2^S(x,t)}{\partial x^2} + \ldots\ldots .  \quad (3)$$

When we apply the Regge theory, we take the form of singlet structure function as $F_2^S(x,t) = Ax^{-\lambda}$, where A is a constant or a function of t, and $\lambda$ is the intercept. Then

$$\frac{\partial F_2^S(x,t)}{\partial x} = A(-\lambda)x^{-\lambda-1} = Ax^{-\lambda}(-\lambda)x^{-1} = (-1)\lambda x^{-1} F_2^S(x,t),$$

$$\frac{\partial^2 F_2^S(x,t)}{\partial x^2} = A(-\lambda)(-\lambda-1)x^{-\lambda-2} = Ax^{-\lambda}(-\lambda)x^{-1} = (-1)^2 \lambda(\lambda+1)x^{-2} F_2^S(x,t),$$

$$\frac{\partial^3 F_2^S(x,t)}{\partial x^3} = (-1)^3 \lambda(\lambda+1)(\lambda+2)x^{-3} F_2^S(x,t) \text{ and so on. So equation (3) becomes}$$

$$F_2^S\left(\frac{x}{\omega}, t\right) = F_2^S(x,t) + \left(\sum_{k=1}^{\infty} u^k\right)(-1)\lambda F_2^S(x,t) + \frac{1}{2}\left(\sum_{k=1}^{\infty} u^k\right)^2 (-1)^2 \lambda(\lambda+1) F_2^S(x,t) + \ldots\ldots .$$

So in the expansion series, we will get terms with alternate positive and negative signs and contribution from $\lambda$ to each term increases. So in this case, it is not possible to truncate this infinite series into finite number of terms by applying boundary condition such as small-x [25] and also this is not a convergent series [22]. So, in solving DGLAP evolution equation applying Regge behaviour of singlet structure functions we cannot apply Taylor series expansion method. Same thing happens for non-singlet structure function also.

Now let us consider the Regge Behaviour of singlet structure function [12] as

$$F_2^S(x, t) = T(t)\, x^{-\lambda_S}, \tag{4}$$

where T(t) is a function of $Q^2$ only and $\lambda_S$ is the intercept. At first we have taken $\lambda_S$ as a constant factor. From equation (4), we get

$$F_2^S\left(\frac{x}{\omega}, t\right) = T(t)\, \omega^{\lambda_S}\, x^{-\lambda_S}. \tag{5}$$

Let us assume for simplicity [26, 27]

$$G(x, t) = K(x) F_2^S(x, t), \tag{6}$$

where K(x) is a parameter to be determined from experimental data and we take the form of K(x) = k, $ax^b$ or $ce^{dx}$, where k, a, b, c and d are constants. Therefore

$$G\left(\frac{x}{\omega}, t\right) = K\left(\frac{x}{\omega}\right) T(t) F_2^S\left(\frac{x}{\omega}, t\right) = K\left(\frac{x}{\omega}\right) T(t)\, \omega^{\lambda_S}\, x^{-\lambda_S}. \tag{7}$$

Putting equations (4) to (7) in equation (1) we arrive at

$$\frac{\partial T(t)}{\partial t} - \frac{T(t)}{t} P(x) = 0, \tag{8}$$

where

$$P(x) = A_f \left[ \{3 + 4\ln(1-x)\} + 2\int_x^1 \frac{d\omega}{1-\omega}\{(1+\omega^2)\omega^{\lambda_S} - 2\} + \frac{3}{2} N_f \int_x^1 \{\omega^2 + (1-\omega)^2\} K\left(\frac{x}{\omega}\right) \omega^{\lambda_S} d\omega \right].$$

Integrating equation (8) we get

$$T(t) = C\, t^{P(x)}, \tag{9}$$

where C is a constant of integration. Therefore equation (4) becomes

$$F_2^S(x, t) = C\, t^{P(x)}\, x^{-\lambda_S}. \tag{10}$$

At $t = t_0$, equation (10) gives

$$F_2^S(x, t_0) = C\, t_0^{P(x)}\, x^{-\lambda_S}. \tag{11}$$

From equations (10) and (11) we get

$$F_2^S(x, t) = F_2^S(x, t_0) \left(\frac{t}{t_0}\right)^{P(x)}, \tag{12}$$

which gives the t-evolution of singlet structure function in LO for $\lambda_S$ = constant.

Again at $x = x_0$, equation (10) gives

$$F_2^S(x_0, t) = C\, t^{P(x_0)}\, x_0^{-\lambda_S}. \tag{13}$$

From equations (10) and (13), we get

$$F_2^S(x, t) = F_2^S(x_0, t)\, t^{\{P(x) - P(x_0)\}} \left(\frac{x}{x_0}\right)^{-\lambda_S}, \tag{14}$$

which gives the x-evolution of singlet structure function at LO for $\lambda_S$ = constant.

Taking the Regge behaviour of non-singlet structure function for $\lambda_{NS}$ = constant as same as that for singlet structure function and pursuing the same procedure we get from equation (2)

$$F_2^{NS}(x, t) = F_2^{NS}(x, t_0) \left(\frac{t}{t_0}\right)^{Q(x)}, \qquad (15)$$

where

$$Q(x) = A_f \left[ \{3 + 4\ln(1-x)\} + 2\int_x^1 \frac{d\omega}{1-\omega}\left\{\left(1+\omega^2\right)\omega^{\lambda_{NS}} - 2\right\} \right]$$

and

$$F_2^{NS}(x, t) = F_2^{NS}(x_0, t) \, t^{\{Q(x)-Q(x_0)\}} \left(\frac{x}{x_0}\right)^{-\lambda_{NS}}. \qquad (16)$$

Equations (15) and (16) give the t and x-evolutions of non-singlet structure functions for $\lambda_{NS}$ = constant respectively.

We have discussed t and x-evolutions of structure functions with the intercept $\lambda_S$ or $\lambda_{NS}$ as constants. Now let us consider the case where the intercept becomes a function of t. In this case, the Regge behaviour for the singlet structure function becomes

$$F_2^S(x,t) = T(t) \cdot x^{-\lambda_s(t)} \qquad (17)$$

and

$$F_2^S\left(\frac{x}{\omega}, t\right) = T(t) \cdot \omega^{\lambda_s(t)} \cdot x^{-\lambda_s(t)}, \qquad (18)$$

and for gluon structure function

$$G\left(\frac{x}{\omega}, t\right) = k\left(\frac{x}{\omega}\right) \cdot F_2^S\left(\frac{x}{\omega}, t\right) = k\left(\frac{x}{\omega}\right) \cdot T(t) \cdot \omega^{\lambda_s(t)} \cdot x^{-\lambda_s(t)}. \qquad (19)$$

Putting equations (17) to (19) in equation (1) and preceeding same as before, we get

$$\frac{\partial T(t)}{\partial t} - T(t)\left[\ln x \, \frac{\partial \lambda_s(t)}{\partial t} + \frac{1}{t} R(x, t)\right] = 0, \qquad (20)$$

where $R(x,t) = A_f \left[ \{3+4\ln(1-x)\} + 2\int_x^1 \frac{d\omega}{1-\omega}\left\{(1+\omega^2)\,\omega^{\lambda_s(t)} - 2\right\} + \frac{3}{2}\cdot N_f \int_x^1 \{\omega^2 + (1-\omega)^2\} k\left(x/\omega\right) \cdot \omega^{\lambda_s(t)} \, d\omega \right]$

.

Integrating equation (20) we get

$$\ln T(t) = \lambda_s(t) \ln x + \int \frac{R(x,t)}{t} dt. \qquad (21)$$

Experimental results show that the $F_2$ structure function data at small-x for $Q^2 \geq 3.5$ GeV$^2$ can be well described by the parameterization [28]

$$F_2 = C \cdot x^{-\lambda(Q^2)}$$ with $\lambda(Q^2) = A \cdot \ln(Q^2/\Lambda^2)$ with A = 0.0481±0.0013±0.0037. Therefore, let us take

$$\lambda_S(t) = A \cdot t, \tag{22}$$

where A is a constant. Taking the limit of integration from $t_0$ to t and putting equation (22) in (21) we can get the form of T(t) as

$$T(t) = T(t_0) \cdot \exp\left(A \cdot (t-t_0) \cdot \ln x + \int_{t_0}^{t} \frac{R(x,t)}{t} dt\right). \tag{23}$$

Now equation (17) gives

$$F_2^S(x, t) = T(t_0) \cdot \exp\left(A \cdot (t - t_0) \cdot \ln x + \int_{t_0}^{t} \frac{R(x, t)}{t} dt\right) \cdot x^{-A \cdot t}. \tag{24}$$

At $t = t_0$, equation (24) becomes

$$F_2^S(x, t_0) = T(t_0) \cdot x^{-A \cdot t_0}. \tag{25}$$

From equations (24) and (25) we get

$$F_2^S(x, t) = F_2^S(x, t_0) \cdot \exp\left(A \cdot (t - t_0) \cdot \ln x + \int_{t_0}^{t} \frac{R(x, t)}{t} dt\right) \cdot x^{-A(t-t_t)}, \tag{26}$$

which gives the t-evolution of singlet structure function in LO for $\lambda_S(t) = A \cdot t$.

To evaluate x-evolution, we take a constant t. So in this case, $\lambda_S(t) = A \cdot t$ becomes constant for each t. So the $\lambda$ value changes only when t value changes. So we can evaluate x-evolution for $\lambda_S(t) = A \cdot t$ as same in equation (14) only by replacing $\lambda_S$ by $\lambda_S(t) = A \cdot t$.

Taking the Regge behaviour for $\lambda = \lambda(t)$ for non-singlet structure function same as that for the singlet structure function and pursuing the same procedure with $\lambda_{NS}(t) = A_1 \cdot t$, where $A_1$ is an another constant, we get from equation (2)

$$F_2^{NS}(x, t) = F_2^{NS}(x, t_0) \cdot \exp\left(A_1 \cdot (t - t_0) \cdot \ln x + \int_{t_0}^{t} \frac{R(x, t)}{t} dt\right) \cdot x^{-A_1(t-t_t)}. \tag{27}$$

Equation (27) gives the t-evolution of non-singlet structure function in LO for $\lambda_{NS}(t) = A_1 \cdot t$ and similar condition of x-evolution of singlet structure function for $\lambda_S(t) = A \cdot t$ applies for the non-singlet case also. So equation (16) gives the x-evolution of non-singlet structure function only after replacing $\lambda_{NS}$ by $\lambda_{NS}(t) = A_1 \cdot t$.

The deuteron and the proton $F_2$ structure functions measured in DIS can be written in terms of singlet and non-singlet quark distribution functions as

$$F_2^d = \frac{5}{9} F_2^S \tag{28}$$

and

$$F_2^p = \frac{3}{18}F_2^{NS} + \frac{5}{18}F_2^{S}. \tag{29}$$

Putting equation (28) in equations (12) and (14) we get

$$F_2^d(x, t) = F_2^d(x, t_0)\left(\frac{t}{t_0}\right)^{P(x)} \tag{30}$$

and

$$F_2^d(x, t) = F_2^d(x_0, t) \, t^{\{P(x)-P(x_0)\}}\left(\frac{x}{x_0}\right)^{-\lambda_d}. \tag{31}$$

Equations (30) and (31) give t and x-evolutions of deuteron structure functions for $\lambda_S = \lambda_d$ = constant. Putting equation (28) in equation (26) we get

$$F_2^d(x, t) = F_2^d(x, t_0) \cdot \exp\left(A.(t-t_0).\ln x + \int_{t_0}^{t}\frac{R(x,t)}{t}dt\right) \cdot x^{-A(t-t_1)}. \tag{32}$$

Equation (32) gives the t-evolution of deuteron structure function for $\lambda_d(t) = A.t$.

And as we have shown that the x-evolution for $\lambda_S(t) = A.t$ can be evaluated with the equation of x-evolution for $\lambda_S$=constant only after replacing $\lambda_S$ by $\lambda_S(t) = A.t$, so putting equation (28) in equation (14) and replacing $\lambda_S$ by $\lambda_d(t) = A.t$, we get

$$F_2^d(x, t) = F_2^d(x_0, t) \, t^{\{P(x)-P(x_0)\}}\left(\frac{x}{x_0}\right)^{-A.t}, \tag{33}$$

which gives the x-evolution of deuteron structure function for $\lambda_d(t) = A.t$.

Putting equations (12) and (15) in equation (29) we get

$$F_2^p(x,t) = \frac{3}{18}F_2^{NS}(x,t_0)\left(\frac{t}{t_0}\right)^{Q(x)} + \frac{5}{18}F_2^{S}(x,t_0)\left(\frac{t}{t_0}\right)^{P(x)}, \tag{34}$$

where Q(x) and P(x) are different functions of x. So it would not be possible to combine the singlet and non-singlet structure functions to a compact form to extract the proton structure function by this method.

**3 Results and discussion**

In the present paper, we compare our results of t and x-evolutions of deuteron structure function in LO from equations (30) to (33) with the NMC small-x medium-$Q^2$ data [16]. Deuteron structure function $F_2^d(x, Q^2)$ measured in the range of $0.75 \leq Q^2 \leq 7$ GeV$^2$, $0.0045 \leq x \leq 0.0175$ and in the range of $9 \leq Q^2 \leq 20$ GeV$^2$, $0.025 \leq x \leq 0.09$ have been used for phenomenological analysis of t and x-evolutions of deuteron structure functions in LO. The evolutions are made for both $\lambda_d$ = constant and $\lambda_d = \lambda_d(Q^2)$. Here we used the QCD cut-off parameter $\Lambda_{\overline{MS}}$ (N$_f$ = 4) = 323 MeV for $\alpha_s(M_z^2) = 0.119\pm$ 0.002 [29]. We compare our results in each of the equations (30) to (33) for $K(x) = k$, $ax^b$, $ce^{dx}$ where k, a, b, c and d are constants. In our work, we found that the value of the structure function remains almost same for a large range of b, $10^{-2} > b > 10^{-7}$ and the values remain constant also for $b > 10^{-2}$. So we choose $b = 0.01$ for our calculation. Similarly the value of the structure function remains almost same for a large range of d, $-1 > d > -10^{-4}$ and the value also remains constant for $d > -1$. So we choose $d = -1$ for our calculation. The value of $\lambda_d$ should be close to 1/2 in quite a broad range of x [8, 30].

In Figure 1 to Figure 3, we compare our results for t-evolution of deuteron structure function in LO from equation (30) for $\lambda_d = 1/2$ with NMC data. The best fit results were found for $2.25 \leq k \leq 2.93$, $2.33 \leq a \leq 3.08$ and $2.34 \leq c \leq 2.98$.

In Figure 4 to Figure 6, taking average values of k, a and c from Figure 1 to Figure 3 respectively as $k = 2.52$, $a = 2.63$, and $c = 2.6$, we find the best fit results for t-evolution of deuteron structure in LO corresponds to of $0.355 \leq \lambda_d \leq 0.615$.

In Figure 7 to Figure 9, we compare our results for t-evolution of deuteron structure function in LO from equation (32) for $\lambda_d(t) = A.t$, where $A = 0.0481$ [29] with NMC data and find the best fit results corresponds to $1.47 \leq k \leq 1.93$, $1.52 \leq a \leq 2$ and $1.56 \leq c \leq 1.97$.

In Figure 10 to Figure 12, taking average values of k, a, and c from Figure 7 to Figure 9 respectively as $k = 1.62$, $a = 1.68$, and $c = 1.69$, we compare our results for t-evolution of deuteron structure function in LO for $\lambda(t) = A.t$ and find the best fit results for $0.01 \leq A \leq 0.065$.

Here in each figure for t-evolution of deuteron structure function, graphs 'a' to 'd', correspond to $0.0045 \leq x \leq 0.0175$. The best fit graphs are shown by dashed, solid, solid and dotted lines respectively. The limiting values in each case are the values corresponding to $x = 0.0045$ and $x = 0.0175$.

In Figure 13, we compare our results for x-evolution of deuteron structure function in LO for $\lambda_d$ = constant and for $k = 0.01$ from equation (31). We find the best fit results correspond to $0.09 \leq \lambda_d \leq 0.22$. Here we have kept fixed the value of k at 0.01 because we find that the value of structure function in this case remains almost same for the large range of $10^{-6} \leq k \leq 10^{-2}$ and the value remains constant for $k > 10^{-2}$.

In Figure 14, we compare our results for x-evolution of deuteron structure function in LO for $\lambda_d(t) = A.t$ and taking $k = 0.01$ from equation (33). We find the best fit results for $0.02 \leq A \leq 0.042$. We compare our results taking $K(x) = ax^b$ and $K(x) = ce^{dx}$ also and found the results same as $K(x) = k = 0.01$ with $a = c = 0.01$.

Here in each figure for x-evolution of deuteron structure function, graphs 'a' to 'd', correspond to $9 \leq Q^2 \leq 20$ GeV$^2$. The best fit graphs are shown by dashed, solid, solid and dotted lines respectively. The limiting values in each case are the values corresponding to $Q^2 = 9$ GeV$^2$ and $Q^2 = 20$ GeV$^2$. From the x-evolution graphs it is clear that the theory gives better results for $\lambda_d(t) = A.t$ and the results will be more accurate for large-$Q^2$.

## 4 Conclusions

In our present work, we have derived the solutions of singlet and non-singlet DGLAP evolution equations in LO at small-x limit applying Regge theory from which we find the t and x-evolutions of deuteron structure function at small-x in LO [30-33]. The LO deuteron structure functions results for t and x-evolutions are compared with NMC small-x ($0.0045 \leq x \leq 0.0175$) and medium-$Q^2$ ($0.75 \leq Q^2 \leq 7$ GeV$^2$) data from which we find the ranges of the intercept $\lambda_d$ of Regge behaviour for deuteron structure function. Taking $\lambda_d$ = constant we find the range as $0.355 \leq \lambda_d \leq 0.615$ for k = 2.52, a = 2.63, and c = 2.6 and taking $\lambda_d(t) = A.t$ we find the range as $0.01 \leq A \leq 0.065$ for k = 1.62, a = 1.68 and c = 1.69. Here we overcome the limitations arose from Taylor series expansion method and we also discuss the situation why it is not possible to apply Regge behaviour and Taylor series expansion method simultaneously. After applying Regge behaviour of structure function DGLAP evolution equations become quite simple to solve.

Our next work is to carry out the above calculations with next-to-leading order (NLO) DGLAP evolution equations and also to compare the results for both LO and NLO cases and to find the possible NLO contributions. We are also planning to apply Regge behaviour in spin-dependent evolution equations to find out various spin-dependent quantities.

*Acknowledgements.* We are grateful to R. Rajkhowa and R. Baishya of Tezpur University, India for important discussions. We are also grateful to Dr. G. A. Ahmed for his help in numerical part of this work. One of us (JKS) is grateful to the University Grants Commission, New Delhi for the financial assistance to this work in the form of a major research project.

## A Appendix

A simple C-programme for the calculation of t-evolution of deuteron structure function in LO for $\lambda_d$ = constant.

```c
//*Variation of singlet structure function with Q2 for k=constant*//
#include<math.h>
#include<stdio.h>
#include<conio.h>
#define N_f  4
#define Λ    0.323
#define k    2.52
#define λ    0.355
#define lm   0.0125
#define Q_0^2  0.75
#define F_0  0.2538
#define div  1000
#define ul   0.9999
#define f(w)(2*((1+(w*w))*pow(w,λ)/(1-w)-2/(1-w)))
        +(1.5*N_f*k*(((2*w*w)-(2*w)+1)*pow(w,λ))))
#define p(x)  0.16*(3+(4*log(1-lm))+u)
#define F_2^d(x,t)  F_0*pow(((log(Q^2)-log(Λ*Λ))/(log(Q_0^2)-log(Λ*Λ))),p(x))
main()
{
int i;
double h,s,sa,sb,u,Q^2;
clrscr();
h=(ul-lm)/div;
s=f(ul)+f(lm);
```

```
for(i=1;i<=div-1;i=i+1)
{
sa=(lm+(i*h));
s=s+(2*(f(sa)));
}
for(i=1;i<=div-1;i=i+2)
{
sb=(lm+(i*h));
s=s+(2*(f(sb)));
}
u=(s*h)/3.0;
printf("\n integral=%lf", u);
printf("\n value of p(x)=%lf"
, p(x));
printf("\n value of $Q^2$: ");
scanf("%lf",  &$Q^2$);
printf("\n value of $F_2^d(x,t)$=%lf"
, $F_2^d(x,t)$);
getch();
return(0);
}
```

## B Appendix

A simple C-programme for the calculation of t-evolution of deuteron structure function in LO for $\lambda_d =$ A.t.

```
//*t-evolution of deuteron structure function in
LO for λ_d=A*t*//
#include<math.h>
#include<stdio.h>
#include<conio.h>
#define N_f   4
#define Λ    0.323
#define A_f   0.16
#define k    1.62
#define div  1000
#define ul   0.9999
#define t  (log(Q^2)-log(Λ*Λ))
#define t_0 (log(Q_0^2)-log(Λ*Λ))
#define λ_d   A*t
#define λ_d^0  A*t_0
#define A    0.015
#define lm   0.008
#define Q_0^2  0.75
#define f    0.2595
#define f1(w) ((2*((1+(w*w))
 *pow(w,λ_d)/(1-w)-2/(1-w)))
 +(1.5*N_f*k*((2*w*w)-(2*w)+1)
 *pow(w,λ_d)))
#define f2(v)  pow(v,-1)
#define f3  A_f*((3+(4*log(1-
 lm)))+(ux))*(ut)
#define p(x,t)  exp(A*(t-t_0)
 *log(lm)+f3)
#define F_2^d(x,t) f*(p(x,t))
 *(pow(lm,(λ_d^0-λ_d)))
main()
{
int  i;
double  hx,ht,sx,sax,sbx,ux,Q^2,st,sat,sbt,ut;
clrscr();
printf("\n value of Q^2: ");
scanf("%lf",  &Q^2);
printf("\n value of λ_d=%lf",λ_d);
hx=(ul-lm)/div;
sx=(f1(ul)+f1(lm));
for(i=1;i<=div-1;i=i+1)
{
sax=(lm+(i*hx));
sx=sx+(2*(f1(sax)));
}
for(i=1;i<=div-1;i=i+2)
{
sbx=(lm+(i*hx));
sx=sx+2*((f1(sbx)));
}
ux=(sx*hx)/3.0;
printf("\n integral ux=%lf", ux);
ht=(t-t_0)/div;
st=(f2(t)+f2(t_0));
for(i=1;i<=div-1;i=i+1)
{
sat=(t_0+(i*ht));
st=st+(2*(f2(sat)));
}
for(i=1;i<=div-1;i=i+2)
{
sbt=(t_0+(i*ht));
st=st+(2*(f2(sbt)));
}
ut=(st*ht)/3.0;
printf("\n integral ut=%lf", ut);
printf("\n F_2^d(x,t)=%lf", F_2^d(x,t));
getch();
return(0);
}
```

## C Appendix

A simple C-programme for the calculation of x-evolution of deuteron structure function.

```c
//*x-evolution of deuteron structure function*//
#include<stdio.h>
#include<math.h>
#include<conio.h>
#define λ_d   0.22
#define t   (log(Q^2)-log(Λ*Λ))
#define Λ    0.323
#define ul   0.9999
#define div  1000
#define N_f  4
#define A    0.042
#define k    0.01
#define Q^2  9
#define x_0  0.09
#define F_0  0.3656
#define f0(w) ((2*((1+(w*w))
 *pow(w,λ_d)/(1-w)-2/(1-w)))
 +(1.5*N_f*c*exp(d*x0/w)*((
 (2*w*w)-(2*w)+1)*pow(w,λ_d))))
#define p0(x_0) (0.16*(3+
 (4*log(1- x0))+(u0)))
#define f(w) ((2*((1+(w*w))
 *pow(w,λ_d)/(1-w)-2/(1-w)))+
 (1.5*N_f*c*exp(d*x/w)*(((2*w*w)-
 (2*w)+1)*pow(w,λ_d))))
#define p(x)   (0.16*(3+
 (4*log(1-x))+(u)))
#define U   pow(t,(p(x)-
 p0(x_0)))
#define F_2^d(x,t) F_0*
 pow(x_0/x,λ_d)*U
main()
{
int i=0;
double  h,s,sa,sb,u,x,h0,s0,sa0,sb0,u0;
clrscr();
h0=(ul-x0)/div;
s0=f0(ul)+f0(x0);
for(i=1;i<=div-1;i=i+1)
{
sa0=(x0+(i*h0));
s0=s0+(2*(f0(sa0)));
}
for(i=1;i<=div-1;i=i+2)
{
sb0=(x0+(i*h0));
s0=s0+(2*(f0(sb0)));
}
u0=(s0*h0)/3.0;
printf("\n integral=%lf", u0);
printf("\n value of p0(x0)=%lf", p0(x0));
printf("\n value of x: ");
scanf("%lf",  &x);
h=(ul-x)/div;
s=f(ul)+f(x);
for(i=1;i<=div-1;i=i+1)
{
sa=(x+(i*h));
s=s+(2*(f(sa)));
}
for(i=1;i<=div-1;i=i+2)
{
sb=(x+(i*h));
s=s+(2*(f(sb)));
}
u=(s*h)/3.0;
printf("\n integral=%lf", u);
printf("\n value of p(x)=%lf", p(x));
printf("\n U=%lf", U);
printf("\n value of F_2^d(x,t)=%lf",F_2^d(x,t));
getch();
return(0);
}
```

**Figure captions**

**Fig. 1.** t-evolution of deuteron structure function in LO for $\lambda_d = 1/2$ and $K(x) = k$ for the representative values of x. Data points at lowest-$Q^2$ values are taken as input to test the evolution equation (30). Here Fig. 1a-d are the best fit graphs of our result shown by dashed, solid, solid and dotted lines respectively. In each figure, the lower limit of the range of k (k = 2.25) values corresponds to dotted line and the upper limit (k = 2.93) corresponds to dashed line.

**Fig. 2.** Same as Fig. 1 for $K(x) = ax^b$ and b = 0.01. Here we observe the range of 'a'.

**Fig. 3.** Same as Fig. 1 for $K(x) = ce^{dx}$ and d = -1. Here we observe the range of 'c'.

**Fig. 4.** Same as Fig. 1 for K(x) = k = 2.52. Here we observe the range of $\lambda_d$ for $K(x) = k$.

**Fig. 5.** Same as Fig. 2 for a = 2.63 and b = 0.01. Here we observe the range of $\lambda_d$ for $K(x) = ax^b$.

**Fig. 6.** Same as Fig. 3 for c = 2.6 and d = -1. Here we observe the range of $\lambda_d$ for $K(x) = ce^{dx}$.

**Fig. 7.** t-evolution of deuteron structure function in LO for $\lambda_d(t) = A.t$ where A = 0.0481 and $K(x) = k$ for the representative values of x. Data points at lowest-$Q^2$ values are taken as input to test the evolution equation (32). The best fit graphs of our result and the limiting values of range of k are shown similarly as in Fig. 1.

**Fig. 8.** Same as Fig. 7 for $K(x) = ax^b$ and b = 0.01. Here we observe the range of 'a'.

**Fig. 9.** Same as Fig. 7 for $K(x) = ce^{dx}$ and d = -1. Here we observe the range of 'c'.

**Fig. 10.** Same as Fig. 7 for K(x) = k = 1.62. Here we observe the range of A for $K(x) = k$.

**Fig. 11.** Same as Fig. 8 for a = 1.68 and b = 0.01. Here we observe the range of A for $K(x) = ax^b$.

**Fig. 12.** Same as Fig. 9 for c = 1.69 and d = -1. Here we observe the range of A for $K(x) = ce^{dx}$.

**Fig. 13.** x-Evolution of deuteron structure function in LO for $\lambda_d$ = constant and K(x) = k = 0.01 for the representative values of $Q^2$. Data points for x values just below 0.1 are taken as input to test the evolution equation (31). The best fit graphs of our result and the limiting values of range of $\lambda_d$ are shown similarly as in Fig. 1.

**Fig. 14.** Same as Fig. 13 for $\lambda_d(t) = A.t$. Here we observe the range of A from equation (33).

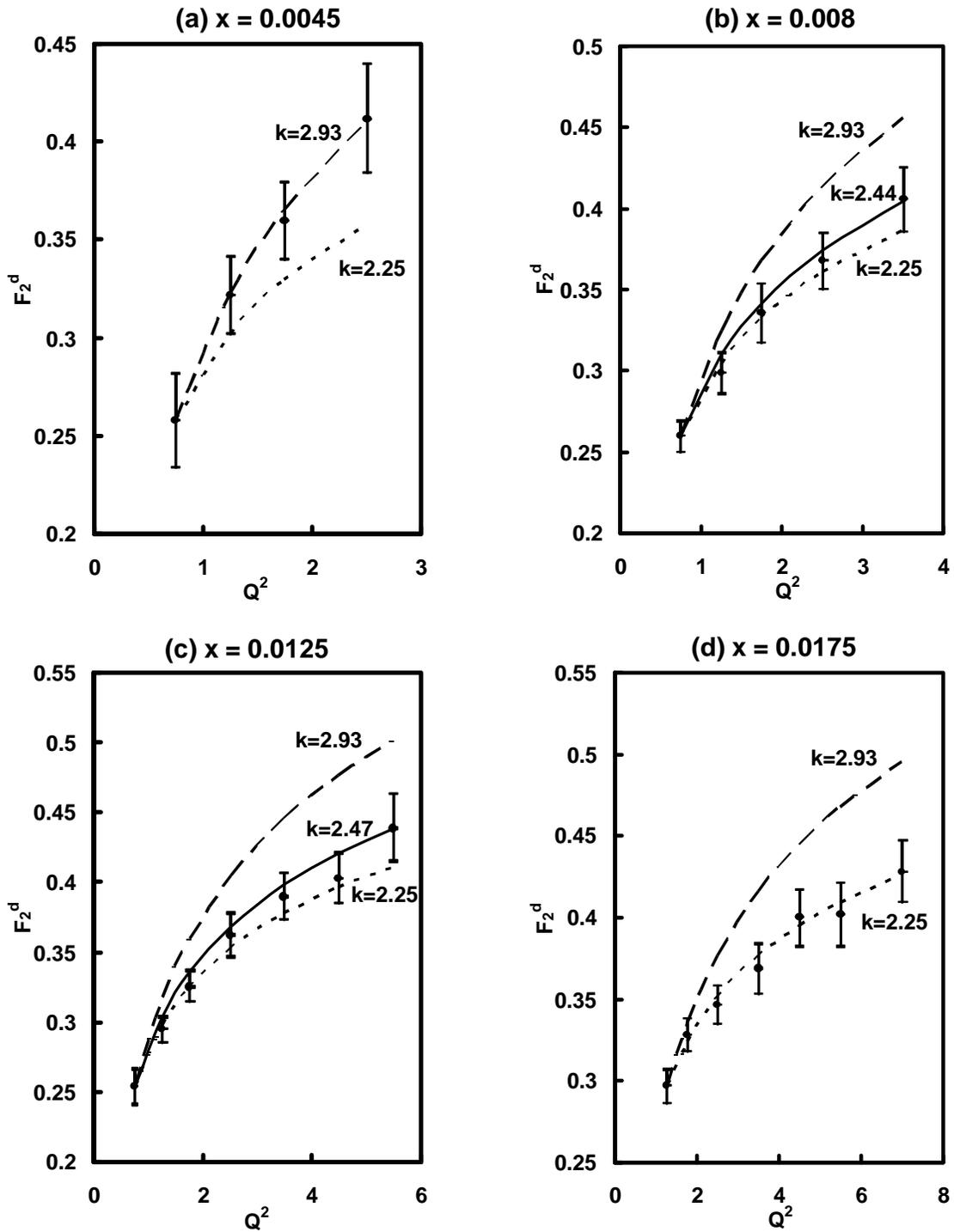

**Fig. 1, k = cont., λ = 0.5**

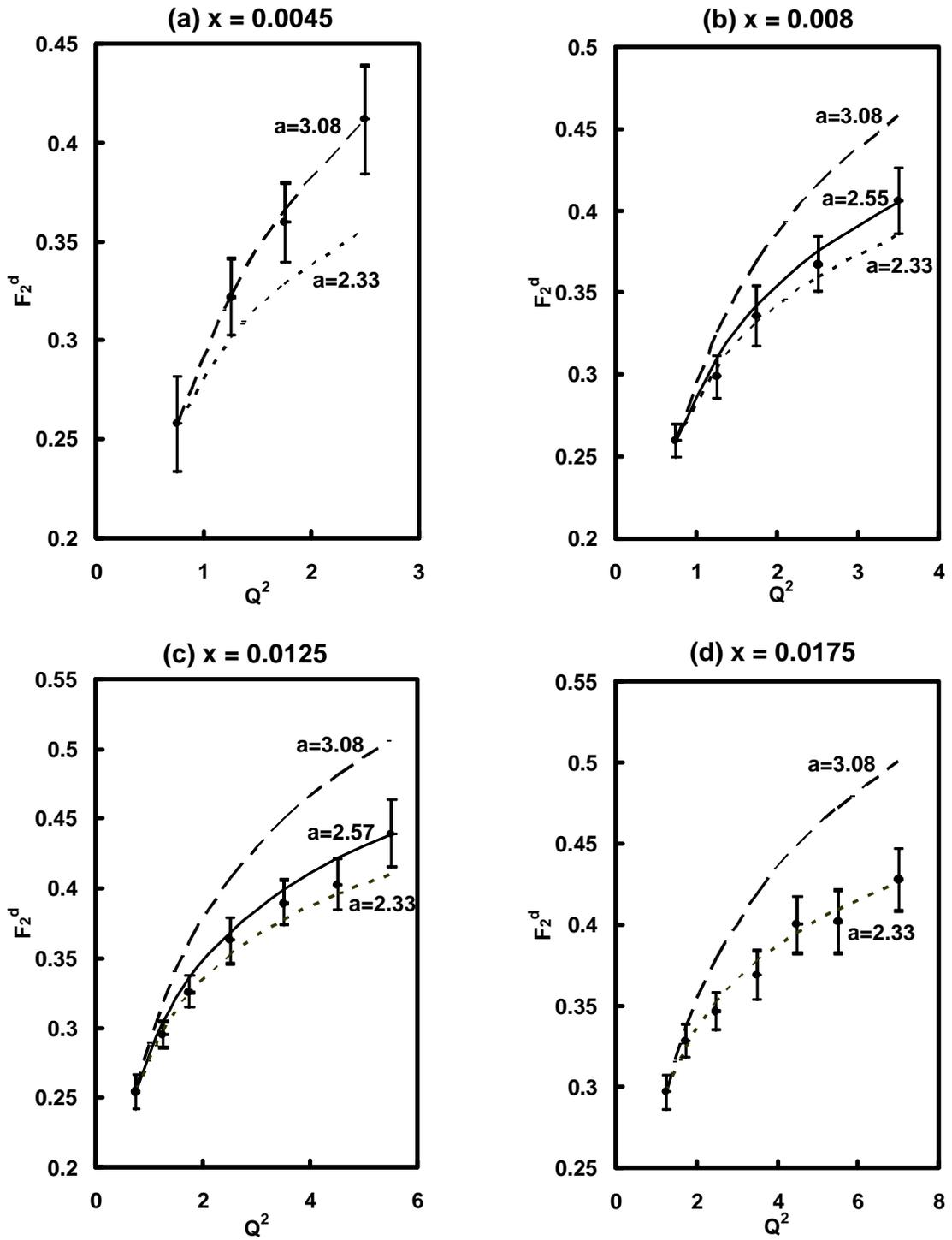

**Fig. 2**, $\lambda = 0.5$, $k = a.x^b$, $b = 0.01$

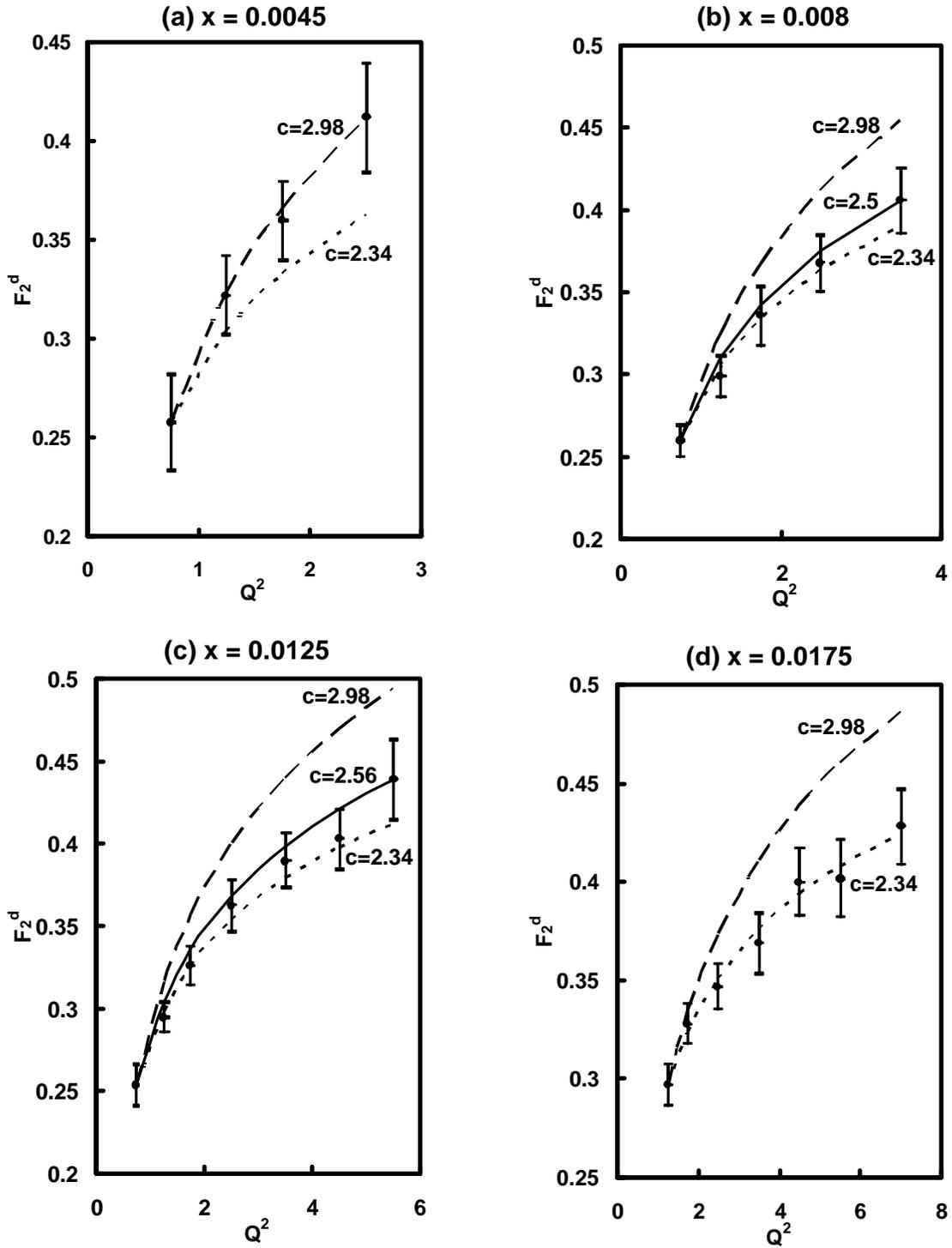

**Fig. 3,** $\lambda = 0.5$, $k = c.e^{d.x}$, $d = -1$

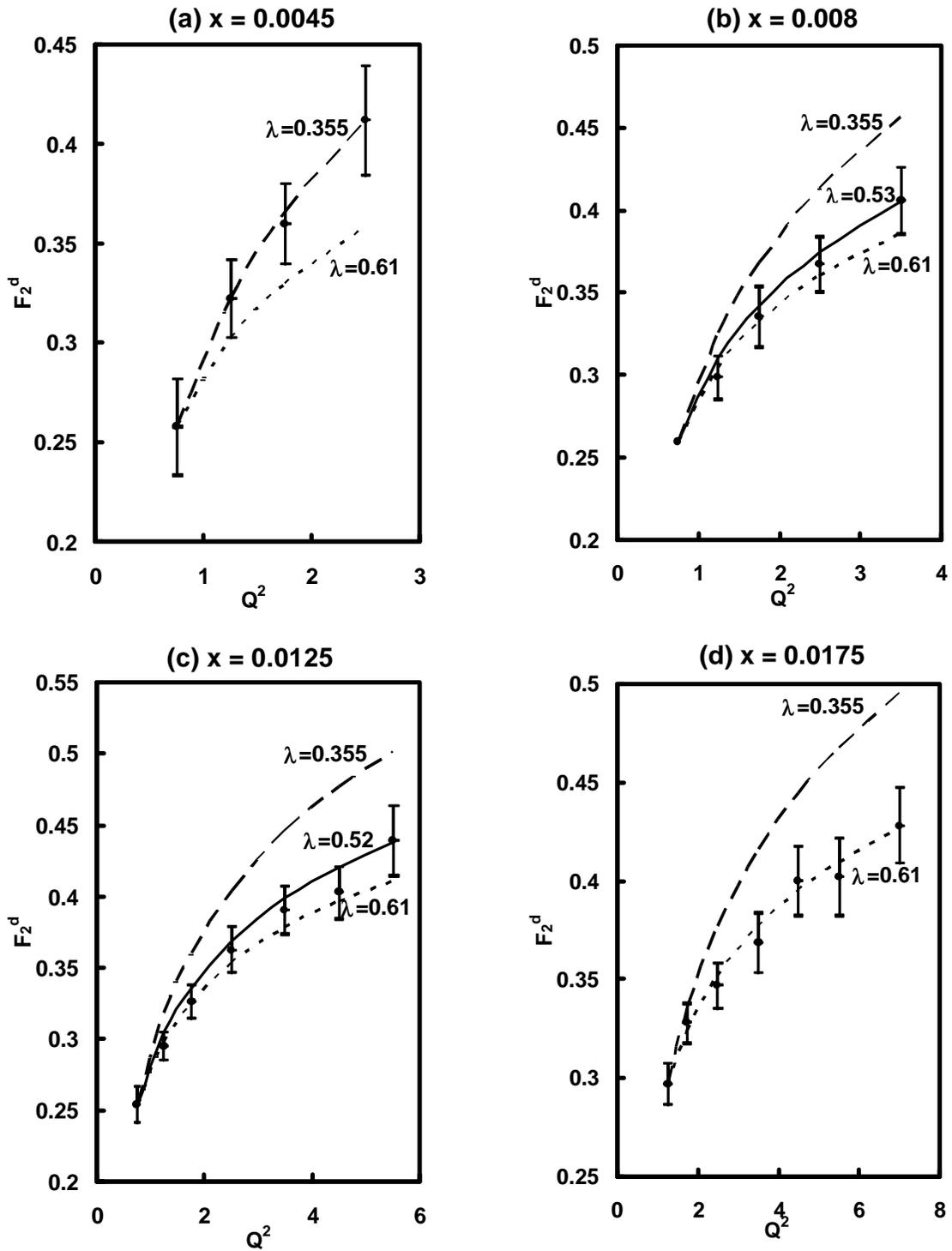

**Fig. 4, k = 2.52**

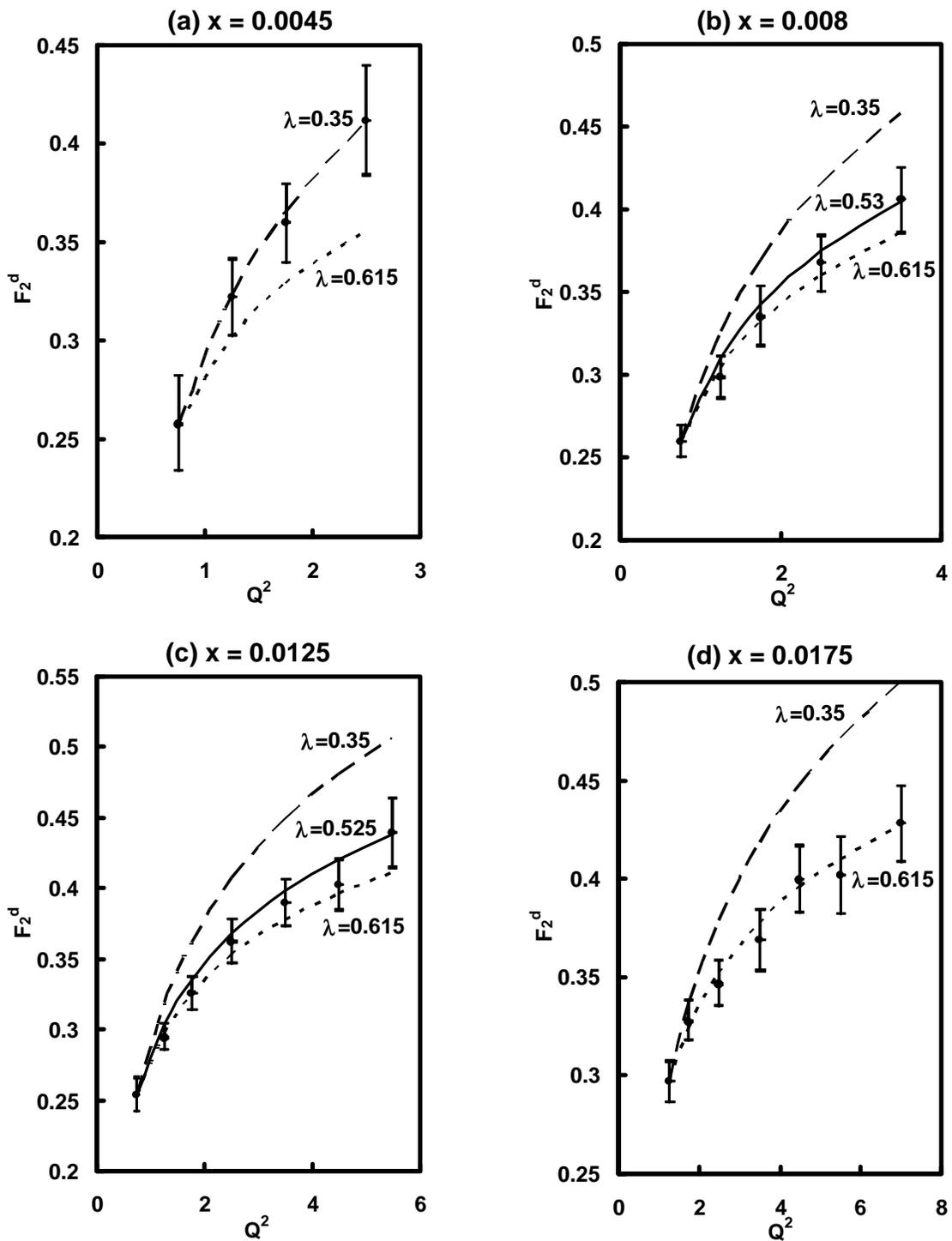

**Fig. 5, k = a.x^b, a = 2.63, b = 0.01**

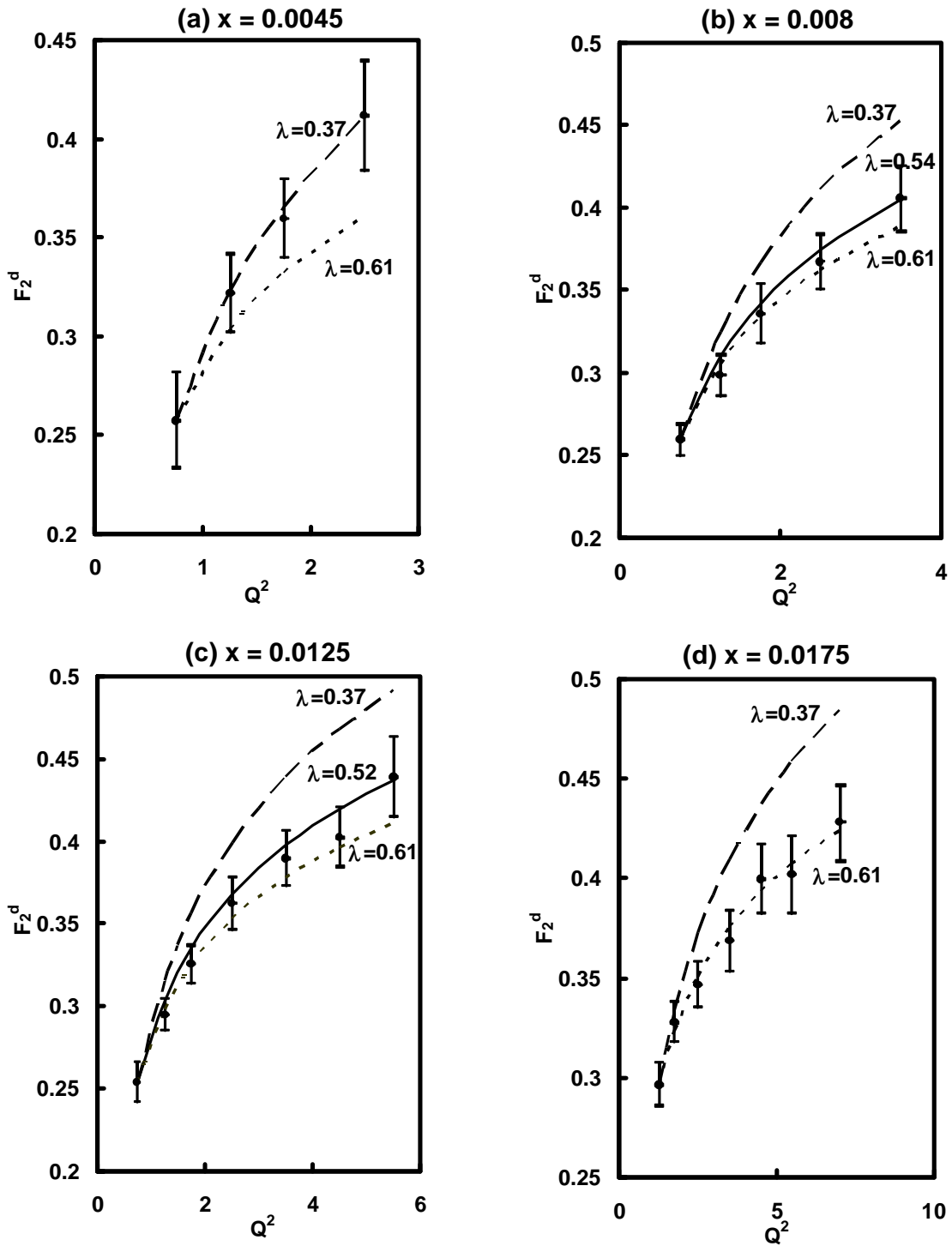

**Fig. 6, $k = c \cdot e^{d \cdot x}$, $c = 2.6$, $d = -1$**

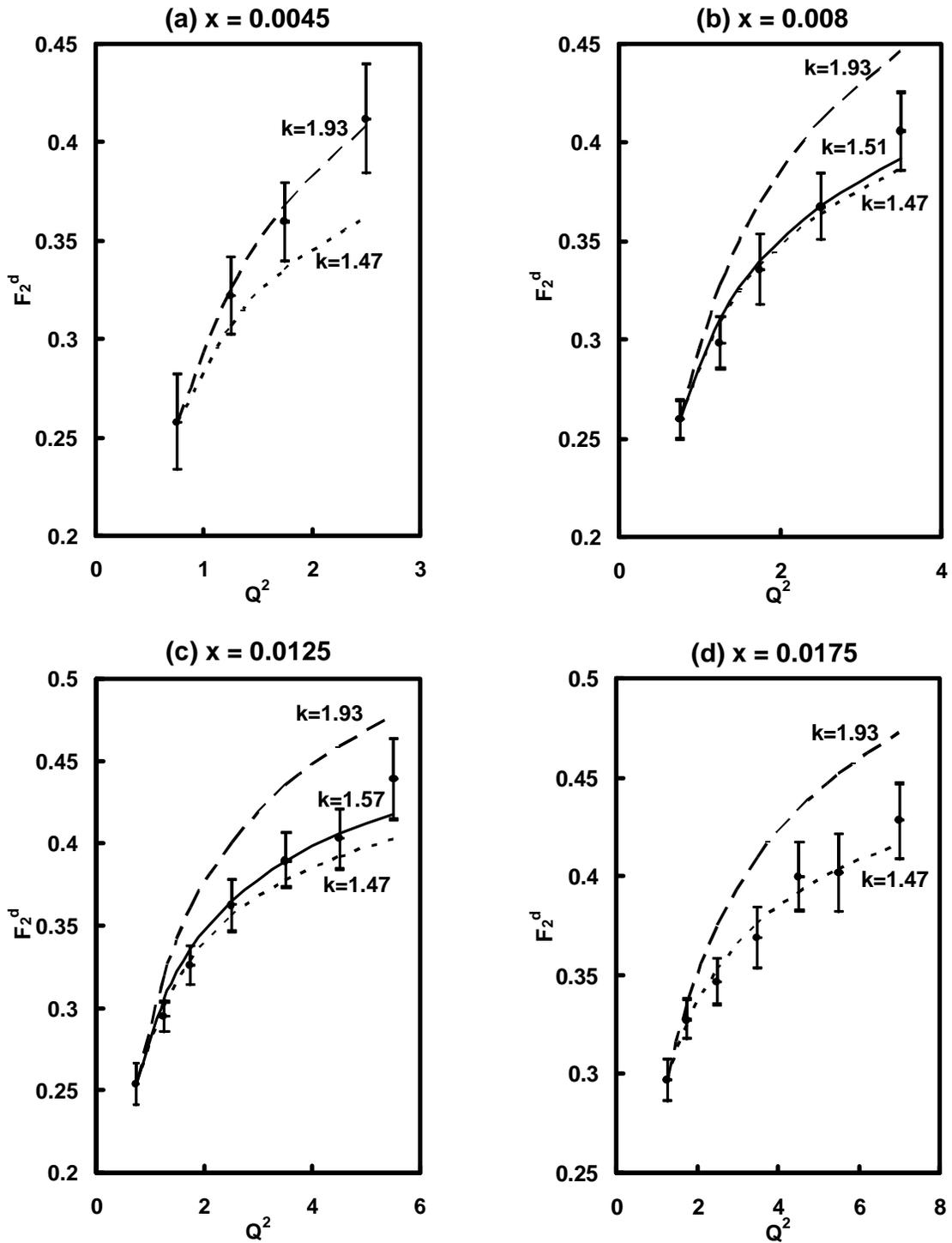

**Fig. 7, A = 0.0481, k = cont., $\lambda$ = A.t**

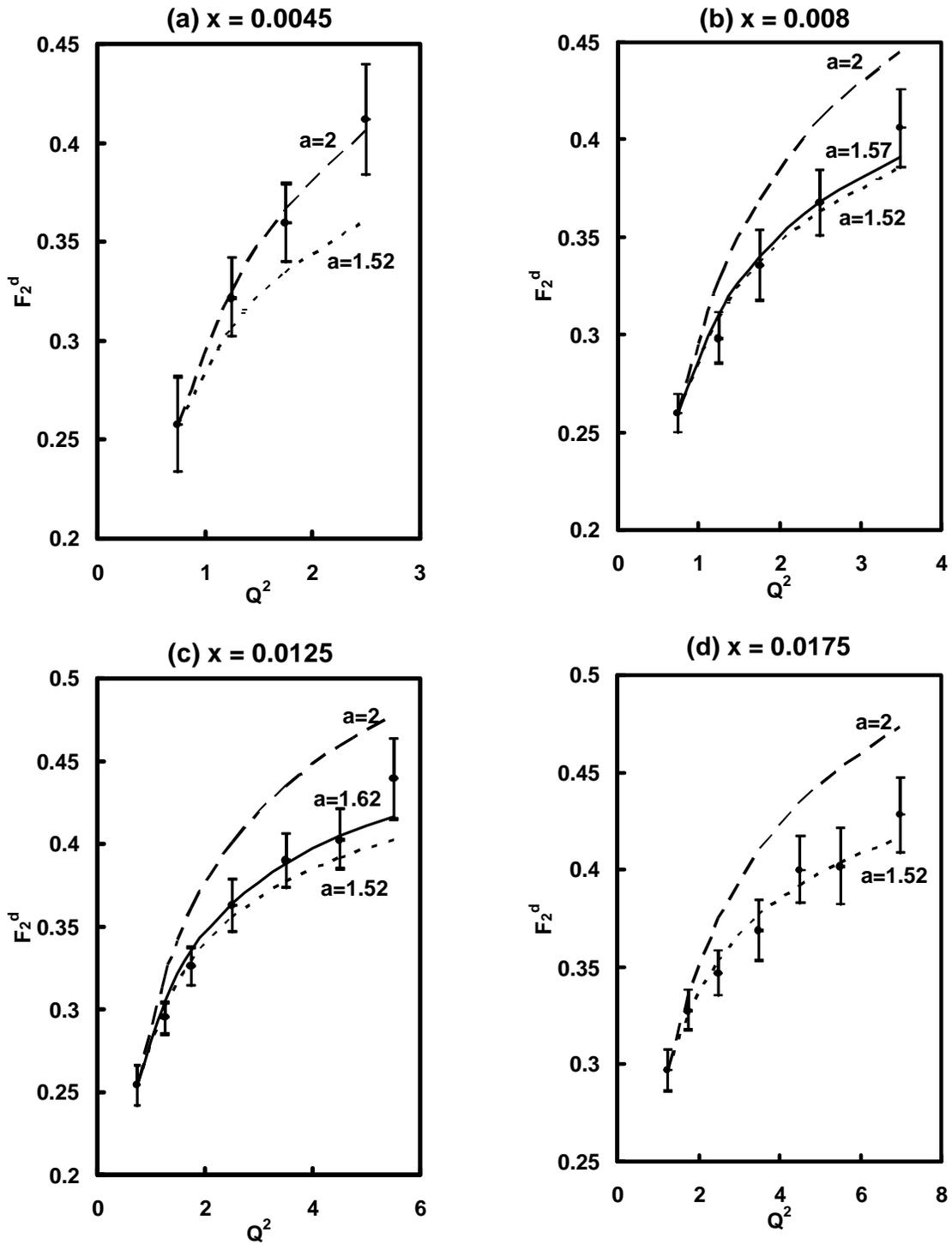

**Fig. 8, k = a.x$^b$, b = 0.01, A = 0.0481, λ = A.t**

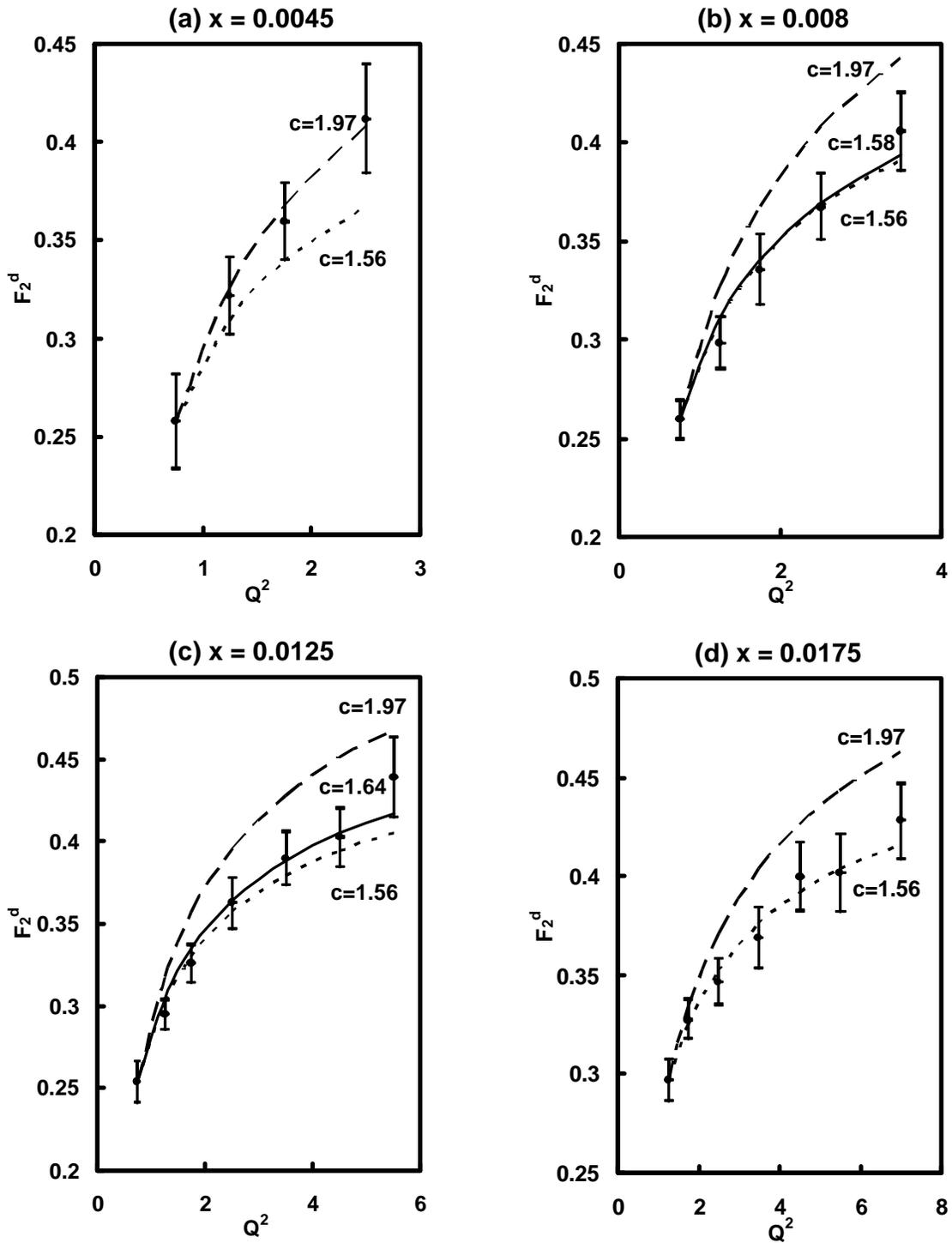

**Fig. 9, k =c.e$^{d.x}$, d = -1, A = 0.0481, λ = A.t**

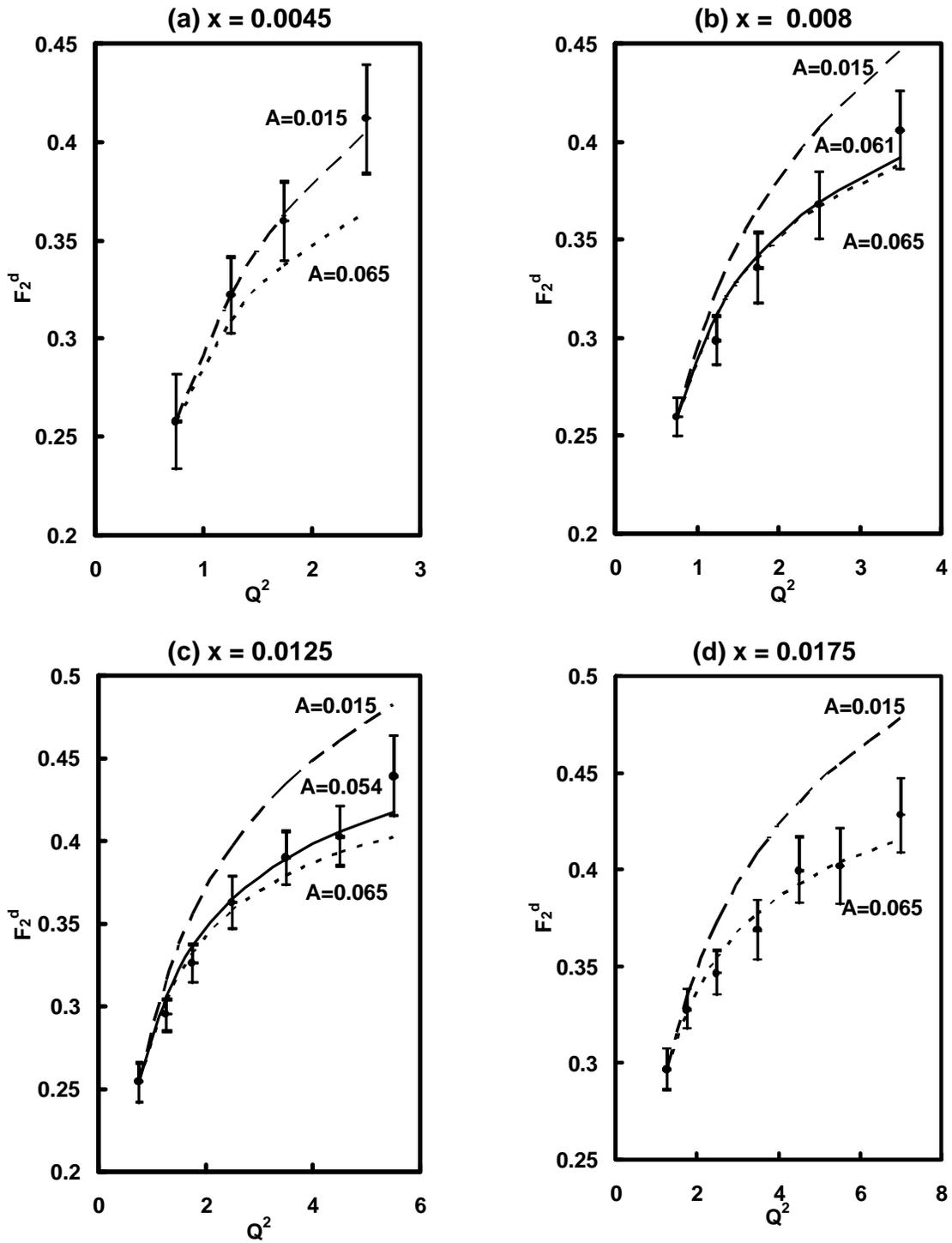

**Fig. 10, k = 1.62, λ = A.t**

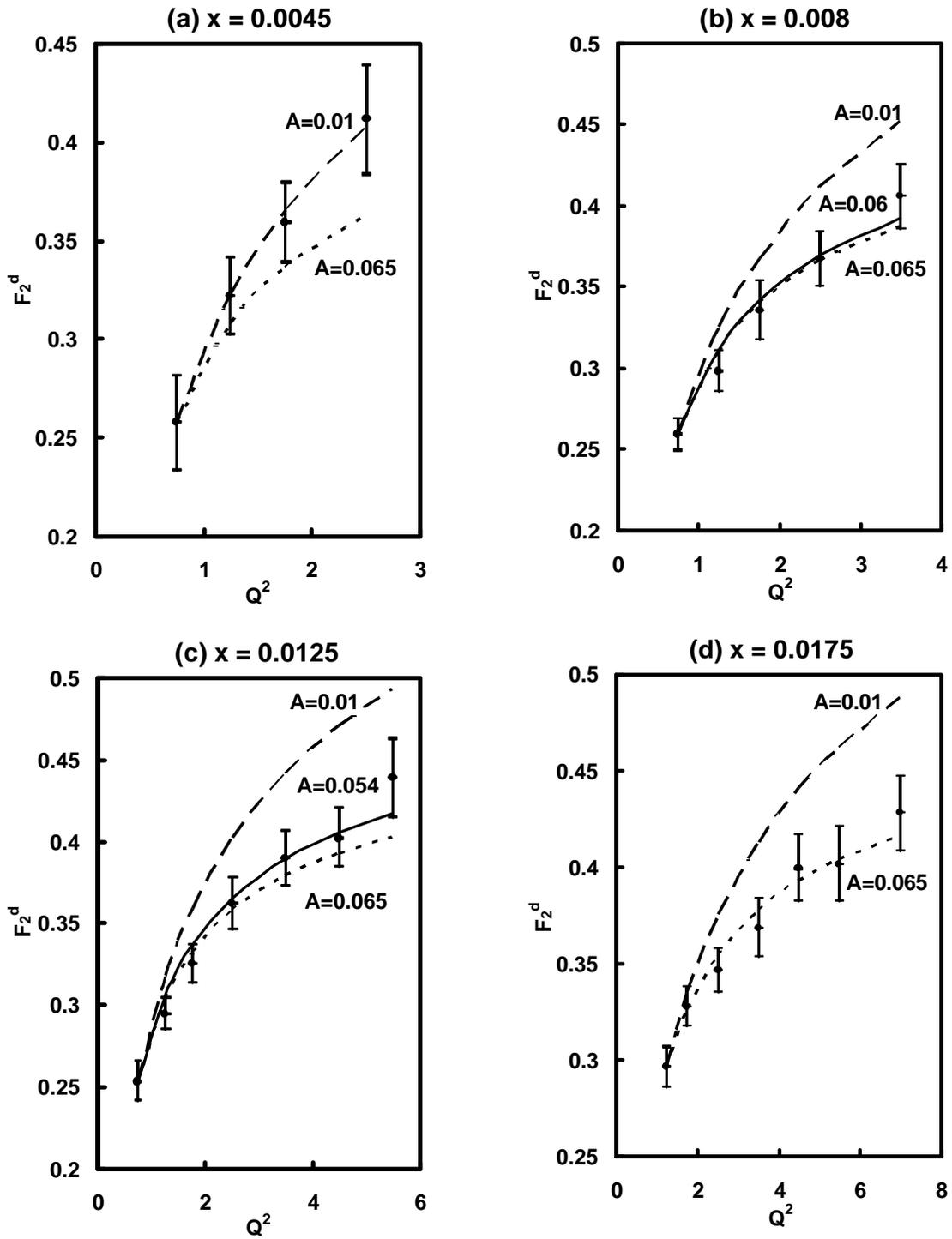

**Fig. 11, k = a.x$^b$, a = 1.68, b = 0.01, λ = A.t**

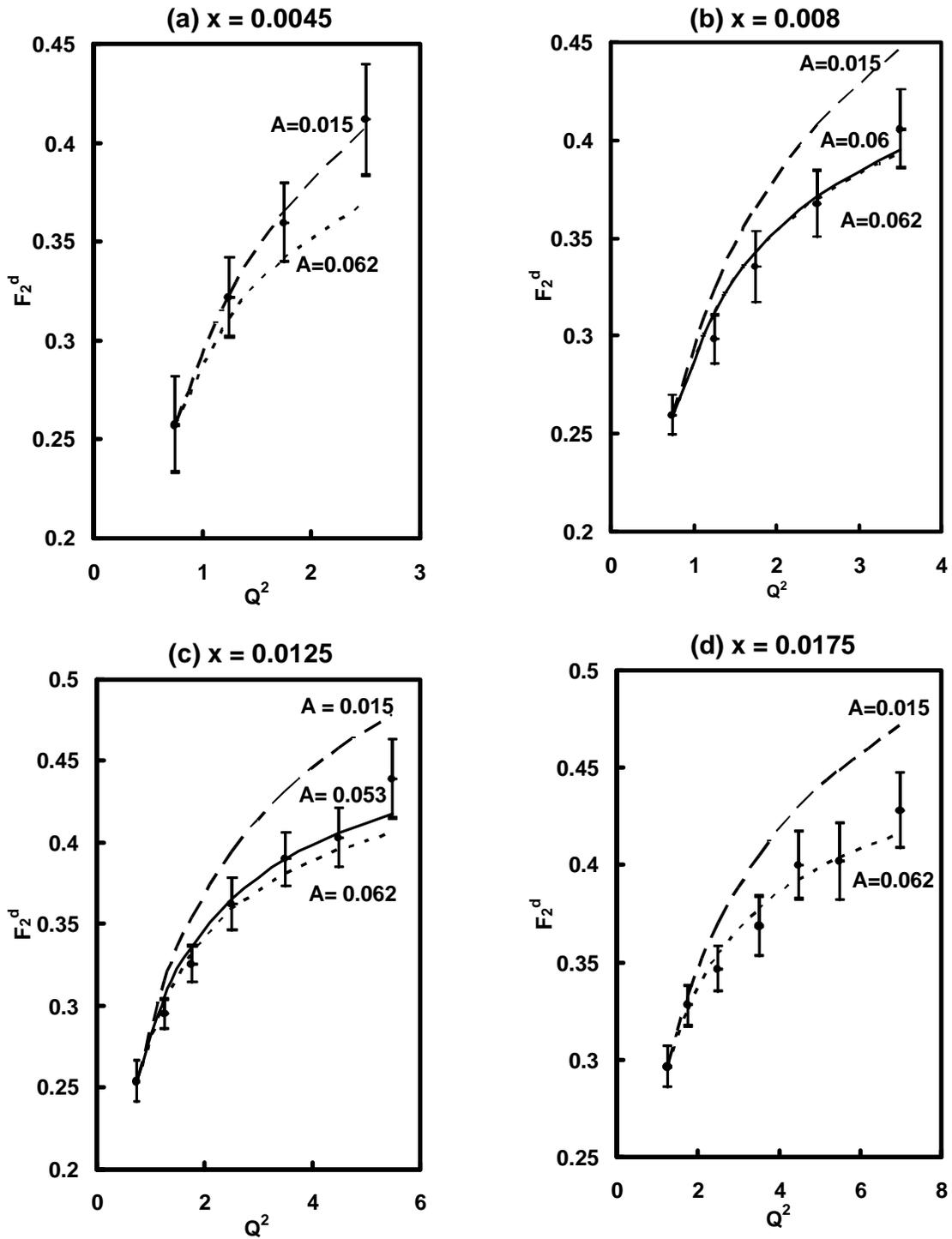

**Fig. 12, $k = c \cdot e^{d \cdot x}$, $c = 1.69$, $d = -1$, $\lambda = A \cdot t$**

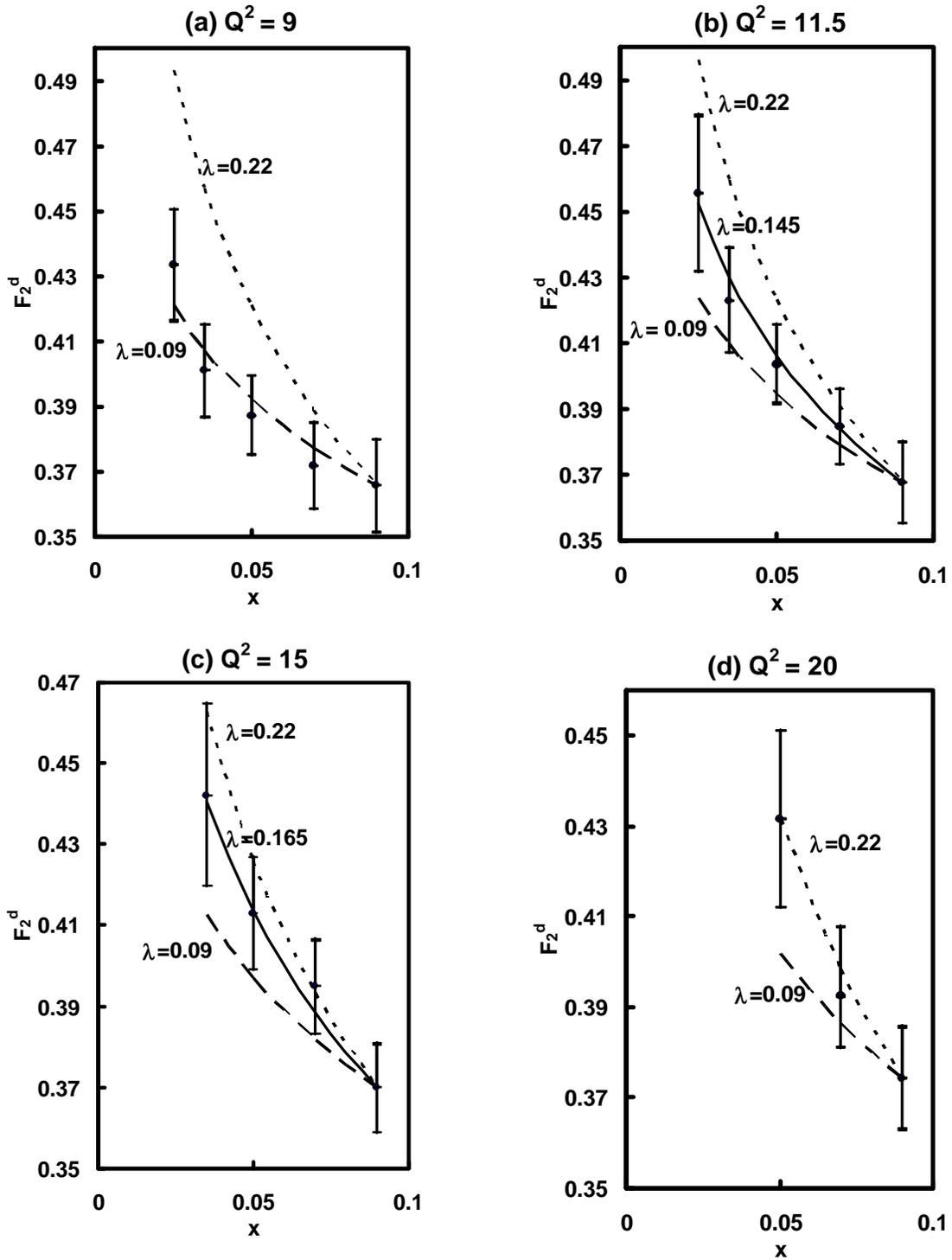

**Fig. 13, k = 0.01**

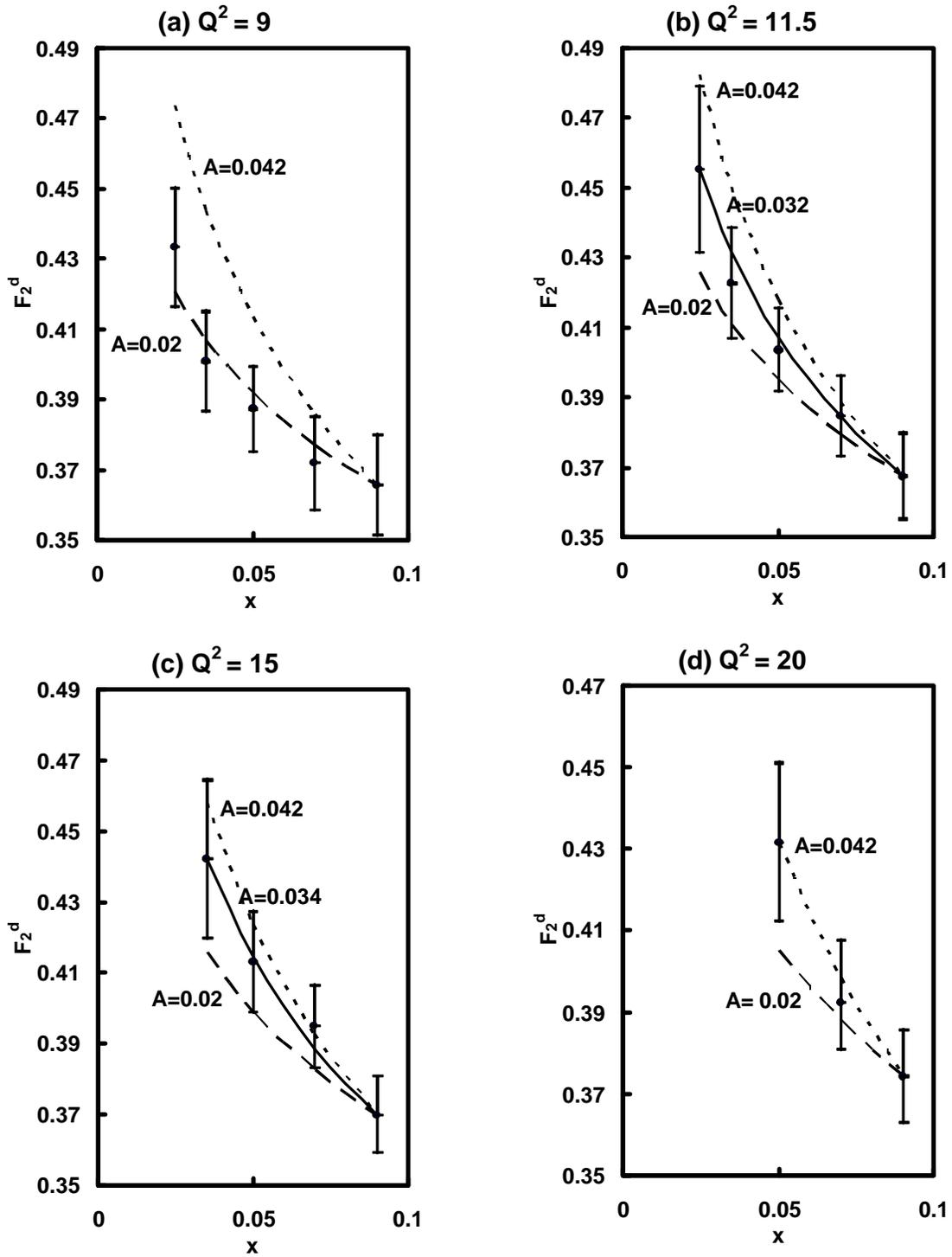

**Fig. 14, k = 0.01, λ = A.t**